\pdfoutput=1

\documentclass[aps,pra,10pt,twocolumn,showpacs,superscriptaddress,nofootinbib]{revtex4}
\usepackage{amsmath}
\usepackage{latexsym}
\usepackage{amssymb}
\usepackage{graphics,epstopdf}

\begin{document}

\title{Stronger steerability criterion for more uncertain continuous
variable systems}

\author{Priyanka Chowdhury}
\email{priyanka@bose.res.in}
\affiliation{S. N. Bose National Centre for Basic Sciences, Block JD, Sector III, Salt Lake, Kolkata 700098, India}

\author{Tanumoy Pramanik}
\email{Pramanik@telecom-paristech.fr}
\affiliation{LTCI, T\'{e}l\'{e}com ParisTech, 23 avenue dItalie, 75214 Paris CEDEX 13, France}

\author{A. S. Majumdar}
\email{archan@bose.res.in}
\affiliation{S. N. Bose National Centre for Basic Sciences, Block JD, Sector III, Salt Lake, Kolkata 700098, India}

\begin{abstract}

We derive a fine-grained uncertainty relation for the measurement of two 
incompatible observables on a single quantum system of continuous variables,
and show that continuous variable systems are more
uncertain than discrete variable systems. Using the derived fine-grained 
uncertainty relation, we formulate stronger steering criterion that  
is able to reveal the steerability of N$00$N states that has hitherto not 
been possible using other criteria. We further obtain a monogamy
relation for our steering inequality which leads to an, in principle,
improved lower bound on the secret key rate of a one-sided device independent
quantum key distribution protocol for continuous variables.


\end{abstract}

\date{\today}

\maketitle

\section{Introduction}

First, Heisenberg in his seminal paper~\cite{Heisenberg} noted that two 
non-commuting observables in quantum mechanics could not be measured with 
arbitrary precision. The uncertainty 
principle introduces a sharp distinction between quantum and classical 
physics. Subsequently, a number of improved uncertainty relations have 
been provided~\cite{GUR, bial, Maassen, Wehner_U, Oppenheim}.  
The presence of uncertainty relations endows quantum mechanics with
significant advantages over classical mechanics for performing different 
information processing tasks. Various versions of uncertainty
relations have been used to detect entanglement~\cite{Ent_Uncer,Wehner_U}, to 
classify mixedness of states~\cite{Purity_Uncer}, to categorize 
different physical theories according to their strength of 
nonlocality~\cite{Oppenheim, NL_FUR}, and to bound information leakage in 
key distribution~\cite{devetak, Q_Memory, 1SDIQKD, tanuprl1, St_Dis_Fur,CV_QKD}. 

Uncertainty relations are linked directly to the ability of quantum states 
to enable steering. The phenomenon of quantum steering~\cite{Schrodinger} 
emerges
from the  EPR paradox~\cite{EPR} that was first formulated for
experimental realization \cite{Reid} based on the  Heisenberg uncertainty 
relation. Later,  steering criteria \cite{Walborn} have been formulated 
using entropic uncertainty relations~\cite{bial,Maassen}, which  
reveal steering in states possessing
higher order correlations \cite{Priyanka_steering}. Experimental
demonstrations of steering have been performed beginning with the 
original demonstration of the EPR paradox~\cite{ou}, as well as using 
different settings~\cite{saunders}  and loophole
free arrangements~\cite{loopsteer}. 
Nonetheless, there 
exist states such as the  N$00$N state which fail to display steering 
using the existing steering criteria for higher values of $N$ in spite of 
violating 
Bell-type inequalities~\cite{bell}. Such a feature calls for further 
improved steerability
conditions since steering lies between 
entanglement  and nonlocality  in the hierarchy~\cite{Wiseman_1} 
of quantum correlations.

The tightest steering inequality in discrete variable systems is 
obtained~\cite{St_Dis_Fur} through the application of the fine-grained 
uncertainty relation (FUR)~\cite{Oppenheim}. Fine-graining makes it 
possible to
distinguish the uncertainty inherent in obtaining
any particular combination of outcomes for different measurements. Application
of fine-graining leads to discrimination between various physical theories
based on the strength of nonlocality \cite{Oppenheim,NL_FUR}. It also
provides an
optimal lower bound of entropic uncertainty in presence of quantum 
memory \cite{tanuprl1}. In the present work we derive a FUR for
continuous variables. Using the derived FUR  we formulate a continuous 
variable fine-grained steering criterion, 
and show that this criterion is monogamous, i.e., Alice's steerability of 
Bob to a particular state chosen randomly from the eigenstates of two 
incompatible observables restricts the control of Bob's system by any 
eavesdropper.  The security of one-sided device independent quantum key 
distribution (1s-DIQKD)
\cite{1SDIQKD} protocols rely on demonstration of steering. Here we
show the possibility of obtaining of an improved lower bound on
the key rate of a 1s-DIQKD protocol using our stronger steering inequality
for continuous variables. 

A comparison of the steering inequality derived here 
with the fine-grained steering inequality
for discrete variables~\cite{St_Dis_Fur} is facilitated by
using the Wigner function for computing various probabilities associated
with continuous variables. There exists an
analogy between the measurement of spin-$1/2$ projectors and
the parity operator, since the measurement outcomes for both are dichotomic.
It is well known~\cite{wigner} that the Wigner function
expressed as an expectation value of a product of displaced
parity operators can be used to derive Bell-CHSH inequalities~\cite{bell}
for continuous variables. Here we use the Wigner function 
formalism~\cite{wigner} to derive a steering inequality for 
continuous variables.
We show that fine-graining leads to a novel manifestation of higher 
uncertainty
 in continuous variable systems, thereby enhancing, in principle, the key 
rate of a
  1s-DIQKD protocol.

\section{FINE-GRAINED UNCERTAINTY RELATION FOR CONTINUOUS VARIABLES}

Let us begin with a brief description of EPR-steering by considering the 
following game~\cite{Wiseman_1}. Alice prepares two systems $A$ and $B$ in the 
state $\rho_{AB}$ and sends the system $B$ to Bob. Alice's task is to convince 
Bob that the prepared state $\rho_{AB}$ is entangled. Bob does not trust Alice,
but he trusts  that he receives a quantum system $B$. He is convinced
only when the correlation between his outcome $b$ for the measurement 
chosen randomly from the set $\mathcal{B}\in\{\beta_1,\:\beta_2\}$ and Alice's 
outcome $a$ for the measurement chosen randomly from the set $\mathcal{A}\in\{\alpha_1,\:\alpha_2\}$ can not be explained by a local hidden state (LHS) 
model, {\it i.e.}, 
\begin{eqnarray}
P(a_{\mathcal{A}},b_{\mathcal{B}}) = \displaystyle\sum_\lambda P(\lambda) P(a_{\mathcal{A}}|\lambda) P_{Q}(b_{\mathcal{B}}|\lambda),
\label{LHS}
\end{eqnarray}
where $P(\lambda)$ is a positive valued distribution over a set of hidden
variables $\lambda$, and  $P_{Q}$ denotes probability of an outcome obtained 
from a quantum measurement. We assume here that Alice knows about Bob's set of 
observables.

Now, from Eq.~(\ref{LHS}) it is easy to derive the relation
\begin{eqnarray}
P(b_{\mathcal{B}}|a_{\mathcal{A}}) \leq \max_{\lambda}[P_Q(b_{\mathcal{B}}|\lambda)]= P_Q(b_{\mathcal{B}}|\lambda_{\max})
\label{U_Bound}
\end{eqnarray}
 using $\sum_i x_i y_i \leq \max_i[x_i] \sum_i y_i$ $\forall~ \{x_i,\: y_i\} \geq 0$. Next, using $\sum_i x_i y_i \geq \min_i[x_i] \sum_i y_i$ $\forall~ \{x_i,\: y_i\} \geq 0$,  we get
\begin{eqnarray}
P(b_{\mathcal{B}}|a_{\mathcal{A}}) \geq \min_\lambda[P_Q(b_{\mathcal{B}}|\lambda)]= P_Q(b_{\mathcal{B}}|\lambda_{\min})~.
\label{L_Bound}
\end{eqnarray}
Combining the relations (\ref{U_Bound}) and (\ref{L_Bound}), the sum 
of conditional probability distributions according to the LHS model is bounded 
by
\begin{eqnarray}
&\min_{\beta_1,\beta_2}[P_Q(b_{\beta_1}|\lambda_{\min}) + P_Q(b_{\beta_2}|\lambda_{\min})] &\nonumber \\
&\leq  P(b_{\beta_1}|a_{\alpha_1}) + P(b_{\beta_2}|a_{\alpha_2})  \leq &\nonumber \\
&\max_{\beta_1,\beta_2} [P_Q(b_{\beta_1}|\lambda_{\max}) + P_Q(b_{\beta_2}|\lambda_{\max})] &. 
\label{ST_Bound}
\end{eqnarray}
We formulate the FUR for continuous variables in order to obtain the bounds 
of the 
algebraic 
inequality~(\ref{ST_Bound}).

The concept of fine-graining in uncertainty relations was first introduced by 
Oppenheim and Wehner~\cite{Oppenheim} to explain the failure of 
quantum theory to exhibit the full non-local strength allowed by no-signaling 
theory. They bound an event (which is defined by the outcomes chosen using 
imposed restrictions or conditions) by its minimum possible uncertainty,
or maximum possible certainty, for two 
incompatible observables. 
For qubit systems a
game~\cite{Oppenheim} is considered in which a binary 
question $q\in\{0,\:1\}$ is 
given randomly to a player Bob who measures $\sigma_z$ $(\sigma_x)$ when 
$q=0~(1)$ is received. Here, the average uncertainty of getting spin up 
outcome (labeled by ``$b=0$") irrespective of the given question
(or indeed, the average certainty) where the average is taken over all possible
 choice of  measurements,   is bounded by
\begin{eqnarray}
\frac{1}{2}-\frac{1}{2\sqrt2} \leq \frac{1}{2} [P(b_{\sigma_z}=0)+P(b_{\sigma_x}=0)] \leq \frac{1}{2}+\frac{1}{2\sqrt2},
\label{FUR_Dis}
\end{eqnarray}
where the equalities occur for maximally certain states. Here they are the 
eigenstates of the observables $(\sigma_z+\sigma_x)/\sqrt{2}$ and 
$(\sigma_z-\sigma_x)/\sqrt{2}$ for the upper and lower bounds, respectively. 
The bounds remain the same for  the spin down 
outcome ($b=1$). Using the uncertainty relation 
(\ref{FUR_Dis}), the inequality (\ref{ST_Bound}) is bounded by
$[1-\frac{1}{\sqrt2},1+\frac{1}{\sqrt2}]$. Hence, in discrete variable systems,
the shared state is steerable if the value of $\frac{1}{2}[P(b_{\sigma_z}=0)+P(b_{\sigma_x}=0)]$ lies outside the above range~\cite{St_Dis_Fur}, where Alice has 
prior knowledge of
Bob's measurement settings.

In continuous variable systems, Bell's inequality is shown to be violated 
using the Wigner function formalism~\cite{wigner}. Fine-graining connects
uncertainty with nonlocality, and hence for a given Bell-CHSH inequality one
can formulate a FUR for a bipartite system~\cite{Oppenheim}. Similar 
considerations hold true for single particle quantum systems, thus making it
possible to construct a FUR for single systems using the 
Wigner distribution representing the average of displaced parity measurement.  
Let us
here label the outcome of even parity measurement by ``$0$". The 
corresponding projection operator is given by
\begin{eqnarray}
\Pi^+ (\beta) = \mathcal{D}(\beta) \left(\sum_{n=0}^{\infty} |2n\rangle\langle 2n|\right) \mathcal{D}^\dagger(\beta),
\label{Op_Even_Parity}
\end{eqnarray} 
where $\mathcal{D}(\beta)~(=\exp(\beta \hat{b}^{\dagger}-\beta^{\ast} \hat{b}),$ 
is the displacement operator with coherent displacement $\beta$, and
$\hat{b}$ and $\hat{b}^{\dagger}$ the annihilation and creation operators, 
respectively.  Similarly, the projection operator corresponding to the odd 
parity measurement outcome labeled  by ``$1$" is given by
\begin{eqnarray}
\Pi^- (\beta) = \mathcal{D}(\beta) \left(\sum_{n=0}^{\infty} |2n+1\rangle\langle 2n+1|\right) \mathcal{D}^\dagger(\beta).
\label{Op_Odd_Parity}
\end{eqnarray}
The observable associated with the Wigner function is given by
$\hat{\mathcal{W}}(\beta)=\Pi^+ (\beta)-\Pi^- (\beta)$
which can be realized using detectors with the capability of distinguishing 
the number of absorbed photons~\cite{wigner}. We will take $\beta$'s to be 
real displacements in the rest of this work.

 The average certainty of the parity measurement outcome $b$ 
over displacements $\alpha$ and $\beta$ is given by 
$\frac{1}{2} [P(b_{\alpha})+P(b_{\beta})]$. As  in the case of discrete 
variables, the average certainty here too is 
bounded by the minimum uncertainty states. 
In continuous variable systems it is well known that the coherent states 
\begin{eqnarray}
|\gamma\rangle=\exp[-\frac{|\gamma|^2}{2}] \sum_{m=0}^{\infty} \frac{\gamma^m}{\sqrt{m!}} |m\rangle
\label{coherent}
\end{eqnarray}
correspond to the minimum uncertainty states in phase-space. Therefore, we 
obtain the fine-grained bounds on $\frac{1}{2} [P(b_{\alpha})+P(b_{\beta})]$ 
using the coherent states.

\begin{figure}[!ht]
\resizebox{7cm}{4cm}{\includegraphics{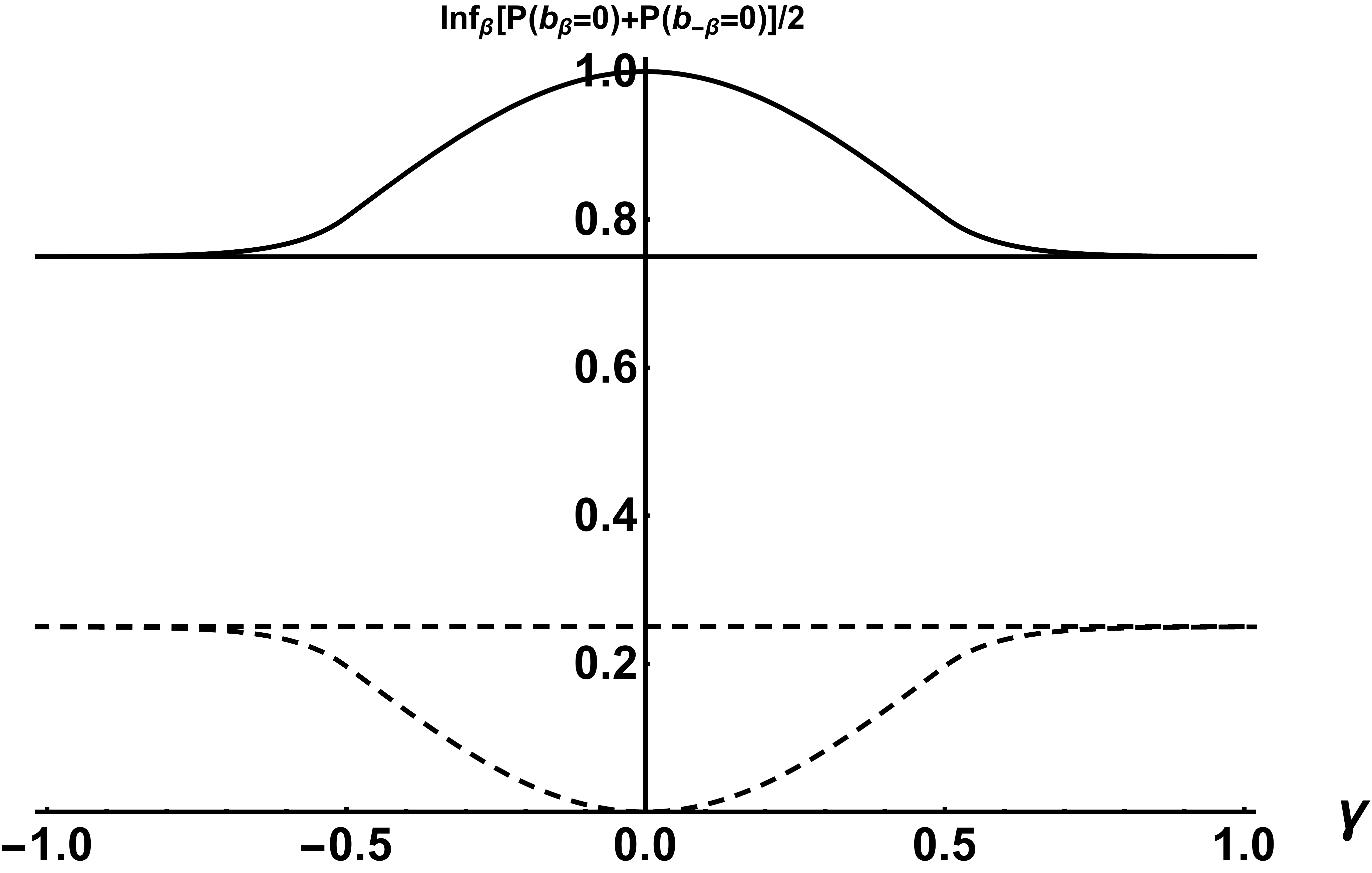}}

\caption{\footnotesize  Plot of $ \frac{1}{2} \inf_{\beta}[P(b_{\beta} = 0)+P(b_{-\beta}=0)]$ with $\gamma$. The solid curve is for $\frac{1}{2} \max_{\beta}[P(b_{\beta} = 0)+P(b_{-\beta}=0)]$, and the dashed curve is for $\frac{1}{2} \min_{\beta}[P(b_{\beta} = 1)+P(b_{-\beta}=1)]$. The solid and dashed lines correspond, respectively, to the upper and lower bounds of certainty in regions of validity of the
FUR~(\ref{FUR_Con}).
\label{compare}
}
\end{figure}

For even parity measurement ($b=0$) at the displacement chosen from 
$\{\alpha,\beta\}$, 
the average certainty becomes
\begin{eqnarray}
\frac{1}{2} [P(b_{\alpha}  &=&  0)+P(b_{\beta}=0)] = \langle\gamma |\frac{\Pi^+(\alpha)+\Pi^+(\beta)}{2}|\gamma\rangle \nonumber \\
&=&\frac{1}{2} \big( \exp[-|\gamma-\alpha|^2] \cosh[|\gamma-\alpha |^2] \nonumber \\
&& +  \exp[-|\gamma-\beta|^2] \cosh[|\gamma-\beta |^2]\big),
\end{eqnarray}
where, similar to the displacements $\beta$ and $\alpha$, we 
choose $\gamma$ also to be real.
 Note here that the condition
$\alpha = \beta \rightarrow \gamma$ needs
to be excluded  in order to ensure that the average certainty of getting
even parity for the zero photon state does not always stay close to $1$,
(this is similar to getting, say, spin up outcome in discrete variables for
spin measurements along directions $\hat{i}$ and $\hat{j}$ when 
$\hat{i} \rightarrow \hat{j}$). For simplification we henceforth
set  $\alpha=-\beta$, and compute the probability distribution
$[P(b_{\beta} = 0)+P(b_{-\beta}=0)]$. The validity of the FUR (\ref{ST_Bound})
is ensured by avoiding the region where
both $\beta \rightarrow 0$ and
$\gamma \rightarrow 0$,
as shown in the Fig.~(\ref{compare})
The  probability distribution $[P(b_{\beta} = 0)+P(b_{-\beta}=0)]$ is bounded by  $[\frac{1}{2},\frac{3}{4}]$, where the maximum occurs for $\beta = \gamma$. 
 Similarly,  the average certainty of  odd parity 
measurements for the displacements $\beta$ and $-\beta$ is bounded by 
\begin{eqnarray}
\frac{1}{2} [P(b_{\beta}  &=&  1)+P(b_{-\beta}=1)] = \langle\gamma |\frac{\Pi^-(\beta)+\Pi^-(-\beta)}{2}|\gamma\rangle \nonumber \\
&=&\frac{1}{2} \big( \exp[-(\gamma-\beta)^2] \sinh[(\gamma-\beta)^2]
\nonumber\\
&& +  \exp[-(\gamma+\beta)^2] \sinh[(\gamma+\beta)^2]\big).
\end{eqnarray}
 Hence, one gets
$\frac{1}{4} \leq \frac{1}{2} [P(b_{\beta} = 1)+P(b_{-\beta}=1)] \leq \frac{1}{2}$,
except at $\gamma\rightarrow 0~\&~\beta\rightarrow 0$ for which 
the 
probability of getting odd counts for the zero photon state approaches zero
(See Fig.~(\ref{compare})). 

Combining the cases of odd and even parities, one  finds that 
the FUR bounds the certainty for the 
measurement of two incompatible continuous variables observables 
 by
\begin{eqnarray}
\frac{1}{4} \leq \frac{1}{2} [P(b_{\beta} )+P(b_{-\beta})] \leq \frac{3}{4}.
\label{FUR_Con}
\end{eqnarray}
In the Fig.~(\ref{compare}), we plot the infimum value of $\frac{1}{2} [P(b_{\beta} )+P(b_{-\beta})]$  with $\gamma$. When $\beta\rightarrow 0$  the 
FUR~(\ref{FUR_Con}) remains valid for  $|\gamma^2| \ge 1$, i.e., for a source 
with at least a single average photon number (we only need to avoid 
the situation when $\gamma\rightarrow 0~\&~\beta\rightarrow 0$ simultaneously).
Note  that 
the range of certainty in discrete variable systems given by Eq.(\ref{FUR_Dis})
\cite{St_Dis_Fur}  is 
higher than that in continuous variable systems. This feature reflecting
more uncertainty  helps to improve the secret 
key rate, since higher uncertainty enables less information to the  
eavesdropper.

Now, using the FUR~(\ref{FUR_Con}), the steering inequality~(\ref{ST_Bound}) becomes
\begin{eqnarray}
\frac{1}{4} \leq  \frac{1}{2}[P(b_{\beta}|a_{\alpha_1}) + P(b_{-\beta}|a_{\alpha_2})] \leq \frac{3}{4}. 
\label{ST_Bound_Con}
\end{eqnarray}
The violation of the inequality~(\ref{ST_Bound_Con}) indicates that the 
measurement correlations are unable to be explained with the help of a LHS 
model, i.e., the given state is  steerable.  Next, we provide an
application of our derived steering inequality  
for N00N states.

\section{STEERABILITY OF NOON STATES}
 
N00N states \cite{lee} are regarded to be of high utility in quantum 
metrology for making
precise interferometric measurements. Such a state is a maximally 
path-entangled two-mode number state of 
continuous
variables,  given by \cite{gsabook}
\begin{eqnarray}
|\text{N00N} \rangle = \frac{1}{\sqrt{2}}(|N,0 \rangle - |0,N \rangle),
\label{NOON}
\end{eqnarray}
where $ N $ is the number of photons either in the first  or 
the second mode. These states have been experimentally realized up to $N=5$
\cite{afek}.  The entanglement of N00N states given in terms of their 
logarithmic negativity is independent of the value of $N$, and 
Bell's inequality is maximally violated for all $N$~\cite{Bell_type_N00N}. 
However, they do not violate the entropic steering 
inequality for $N\geq2$~\cite{Priyanka_steering}. We  show now that 
such states are steerable for $N\geq 2$ using our derived steering 
inequality.

\begin{figure}[!ht]
\resizebox{9cm}{6cm}{\includegraphics{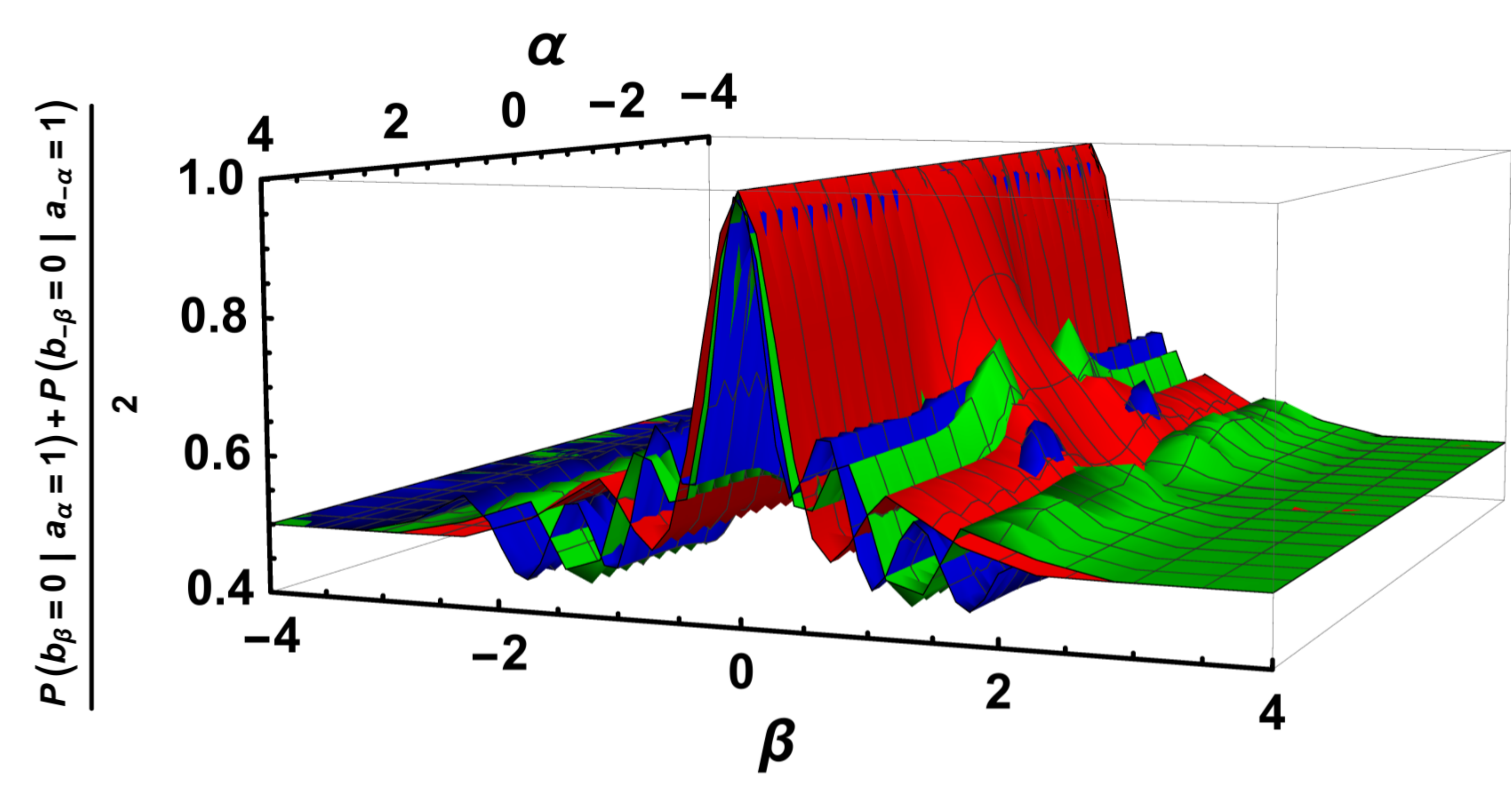}}
\caption{\footnotesize Coloronline. The variation of $\frac{1}{2}[P(b_{\beta}=0|a_{\alpha}=1) + P(b_{-\beta}=0|a_{-\alpha}=1)] $ with respect to $\beta$  and $\alpha$ for three different values of $N$.  (i.) The red colored curve corresponds to the value $N=2$; (iii) the green colored curve is for $N=4$; and the blue colored curve is for $N=6$.
\label{Plot_1}
}
\end{figure}

Considering the cases for $N$ even and odd separately, we find that for
the former
 the maximum violation of the 
inequality~(\ref{ST_Bound_Con}) on the upper side occurs when $\frac{1}{2}[P(b_{\beta}|a_{\alpha_1}) + P(b_{-\beta}|a_{\alpha_2})]=1$ for the choices of the parameters
given by $\{b=0, a=0\}$, $\{b=0,a=1\}$. Similarly, the maximum violation on
the lower side occurs when $\frac{1}{2}[P(b_{\beta}|a_{\alpha_1}) + P(b_{-\beta}|a_{\alpha_2})]=0$ for the choices of the parameters give by $\{b=1, a=0\}$, $\{b=1,a=1\}$. When $N$ is odd, the maximum violations on the upper side are $1$ for 
the choices $\{b=0, a=1\}$, $\{b=1,a=0\}$, and on the lower side are $0$ for 
the choices $\{b=0, a=0\}$, $\{b=1,a=1\}$. In the Fig.~(\ref{Plot_1}), we 
 plot of the quantity $\frac{1}{2}[P(b_{\beta}=0|a_{\alpha}=1) + P(b_{-\beta}=0|a_{-\alpha}=1)] $ versus $\beta$ and $\alpha$  for $N=2,~4,~\text{and~}6$. One sees that the violation of the steering inequality
occurs maximally for $N\geq2$ in the region $|\beta |\rightarrow 0$. 
The condition for validity of our FUR, {\it viz.}, 
 $|\gamma| \ge 1$ when $\beta \rightarrow 0$, is 
ensured since the average photon number ($N/2$ here) is greater than $1$.

\section{SECURITY OF KEY GENERATION}

Steering finds direct applicability in demonstration of security of quantum 
key
distribution (QKD).  The goal of any QKD protocol is to generate a key 
string between two distant parties (say, Alice and Bob) such that it
remains secret from an eavesdropper (say, Charlie). 
In the first QKD 
protocol (BB84) proposed by Bennett and Brassard~\cite{BB84}, security is 
based on 
the uncertainty of the outcome of incompatible spin measurements chosen 
randomly along $x$- and $z$-directions.  The security of standard QKD 
protocols are based on certain idealistic assumptions \cite{QKD_Rev} that
 may be minimized in the so-called device independent QKD
(DIQKD) \cite{diqkd} where it is no longer required to fully trust the 
devices used by Alice and Bob. However, practical and loophole free
implementations of DIQKD protocols are difficult since they require
demonstration of nonlocality.
On the other hand, 1s-DIQKD
\cite{1SDIQKD} protocols which  are intermediate between 
standard QKD and DIQKD
protocols, rely on demonstration of quantum 
steering for their security.
Continuous variable  1s-DIQKD 
protocols have attracted attention in recent
times since they  are reasonably robust to losses and practically more 
feasible
compared to their discrete variable counterparts~\cite{walk}.

Monogamy of non-local correlations certifies the security of 
QKD~\cite{Mono_QKD}. To develop a monogamy relation associated with the upper 
bound of our steering inequality~(\ref{ST_Bound_Con}), let us consider that 
three parties Alice, Bob and Charlie share a tripartite state $\rho_{ABC}$ 
for which the inequality
\begin{eqnarray}
\frac{1}{2}\left(\Sigma_{BA}+\Sigma_{BC}\right) \leq \frac{3}{2}
\label{Mono_1}
\end{eqnarray}
is satisfied, where $\Sigma_{BA}=P(b_{\beta_1}|a_{\alpha_1}) + P(b_{\beta_2}|a_{\alpha_2})$ and $\Sigma_{BC}= P(b_{\beta_1}|c_{\gamma_1}) + P(b_{\beta_2}|c_{\gamma_2})$, 
and $c$ is 
Charlie's outcome for the measurement chosen from the set $\{\gamma_1,\gamma_2\}$. 
The proof comes from contradiction. Let
$\frac{1}{2}\left(\Sigma_{BA}+\Sigma_{BC}\right) > \frac{3}{2}$.
The above sum can be written as the sum of $[P(b_{\beta_1}|a_{\alpha_1}) + P(b_{\beta_2}|c_{\gamma_2})]$ and $[P(b_{\beta_1}|c_{\gamma_1}) + P(b_{\beta_2}|a_{\alpha_2})]$. 
As the 
average of the above two terms is greater than $3/2$, one of the terms, say, 
the first is greater than $3/2$. Then, it is possible to find a conditional 
Bob's state for which $[P(b_{\beta_1})+P(b_{\beta_2})] > 3/2$, which contradicts 
the FUR given by inequality~(\ref{FUR_Con}). Similarly, using the lower bound 
of our steering inequality~(\ref{ST_Bound_Con}), one can obtain
$\frac{1}{2} [ \Sigma_{BA}+\Sigma_{BC} ] \geq \frac{1}{2}$,
which together with the relation (\ref{Mono_1}) gives
\begin{eqnarray}
\frac{1}{2} \leq \frac{1}{2} [ \Sigma_{BA}+\Sigma_{BC} ] \leq \frac{3}{2}.
\end{eqnarray}

The monogamy relation~(\ref{Mono_1}) is used  to bound 
the secret key rate in a 1s-DIQKD  protocol.  The lower 
bound of the secret key rate under individual attack is given by~\cite{CsiszarKorner}
$r\geq \mathcal{I}(B:A)-\mathcal{I}(B:C)$,
where $\mathcal{I}$ is the mutual information. Suppose that the upper 
bound of the steering inequality~(\ref{ST_Bound_Con})  is violated by an 
amount $\delta$, i.e., $\frac{1}{2}[P(b_{\beta_1}|a_{\alpha_1}) + P(b_{\beta_2}|a_{\alpha_2})]= \frac{3}{4} + \delta$, where $0<\delta\leq \frac{1}{4}$. Then, the 
monogamy 
relation (\ref{Mono_1}) implies that $\frac{1}{2}[P(b_{\beta_1}|c_{\gamma_1}) + P(b_{\beta_2}|c_{\gamma_2})]\leq 0.75-\delta$. Hence, the lower bound of the key rate becomes
\begin{eqnarray}
r \geq \log_2\left[ \frac{0.75+\delta}{0.75-\delta}\right],
\end{eqnarray}
where the logarithm of base $2$ is taken since the secret key rate is expressed 
in the units of bits per shared state.
For the maximally entangled  N00N state for which  
$\delta=1/4$, the steering inequality~(\ref{ST_Bound_Con}) is maximally 
violated, making the lower bound  of the secret key rate  
unity. One may note here that in comparison the lower bound 
for discrete variables is $1/2$~\cite{St_Dis_Fur}. Hence, the use
of continuous variable systems in QKD offers more security in principle.

\section{CONCLUSIONS}

To summarize, in the present work we first derive a fine-grained uncertainty 
relation (FUR) for continuous variable systems with the help of an operational 
interpretation of the Wigner function~\cite{wigner}. The FUR provides a 
manifestation of higher uncertainty in 
continuous variable systems than in discrete variable systems, since in 
the former the certainty is confined to a lower range bounded by $[1/4,3/4]$, 
compared to
$[1/2-1/(2\sqrt{2}),1/2+1/(2\sqrt{2})]$ for the latter. The increment 
of uncertainty in continuous variable systems restricts the amount of 
information leakage  to the eavesdropper, making them more secure 
in principle, than discrete variable systems. Our  steering inequality 
is hence also stronger than that of discrete variable systems, as in the 
latter,  
Bob is convinced of the prepared state being entangled only when the 
average of the conditional probabilities is 
larger than $1/2+1/(2\sqrt{2})$~\cite{St_Dis_Fur}, whereas, in continuous 
variable systems it is $3/4$.
Further, we show that our steering inequality is capable of detecting 
maximal steerability by N00N states for 
$N\geq2$, thereby circumventing a drawback of 
the entropic steering inequality which is not violated by the N00N state
for $N\geq2$~\cite{Priyanka_steering}. 

With the help of a  derived  
monogamy relation corresponding to our steering inequality, we bound the 
key rate in the 1s-DIQKD protocol  secured under individual 
attacks. 
The relation of the Wigner
function with displaced parity operators~\cite{wigner} 
facilitates comparison of
the key rates in continuous and discrete variables. The lower bound of the 
secret key rate is unity for the shared maximally entangled state of 
continuous
variables, which is  double that for discrete 
variables~\cite{St_Dis_Fur} even when Alice knows Bob's set of observables 
before preparation of the state. Knowing Bob's set of observables 
does not help Alice to cheat here, whereas it is indeed helpful in discrete 
variable systems~\cite{St_Dis_Fur}.
The quasi-probability distributions contained in our steering 
inequality  may be reconstructed experimentally by homodyne 
detection techniques that are currently realizable with high 
efficiency~\cite{Homodyne_effi}.  Recent experiments~\cite{Expt_Bell_Con} 
have indeed confirmed  Bell violation in continuous variable systems
using similar techniques. It should thus be feasible to experimentally verify 
our steering inequality. Further analysis of more general security attacks,
as well as consideration of decoherence effects would be needed to assess 
the practical viability of such key generation protocols.

{\emph Acknowledgements:}  A.S.M.  acknowledges support from the project SR/S2/LOP-08/2013 of DST, India. T.P acknowledges financial support from ANR retour des post-doctorants NLQCC (ANR-12-PDOC-0022- 01).

\end{document}